\input harvmac
\overfullrule=0pt
\parindent 25pt
\tolerance=10000
\input epsf

\newcount\figno
\figno=0
\def\fig#1#2#3{
\par\begingroup\parindent=0pt\leftskip=1cm\rightskip=1cm\parindent=0pt
\baselineskip=11pt
\global\advance\figno by 1
\midinsert
\epsfxsize=#3
\centerline{\epsfbox{#2}}
\vskip 12pt
{\bf Fig.\ \the\figno: } #1\par
\endinsert\endgroup\par
}
\def\figlabel#1{\xdef#1{\the\figno}}
\def\encadremath#1{\vbox{\hrule\hbox{\vrule\kern8pt\vbox{\kern8pt
\hbox{$\displaystyle #1$}\kern8pt}
\kern8pt\vrule}\hrule}}

\font\cmss=cmss10
\font\cmsss=cmss10 at 7pt

\def\inbar{\vrule height1.5ex width.4pt depth0pt}

 \def\frac#1#2{{#1\over #2}}

 \def\s{\sqrt}

 \def\de{\partial}

 \def\f {\frac}
 \def\ti{\tilde}
 \def\ap{\alpha}

 \def\la{\langle}
 \def\lb{\rangle}

\def\IC{{\relax\,\hbox{$\inbar\kern-.3em{\rm C}$}}}
\def\IZ{\relax\ifmmode\mathchoice
{\hbox{\cmss Z\kern-.4em Z}}{\hbox{\cmss Z\kern-.4em Z}}
{\lower.9pt\hbox{\cmsss Z\kern-.4em Z}} {\lower1.2pt\hbox{\cmsss
Z\kern-.4em Z}}\else{\cmss Z\kern-.4em Z}\fi}
\def\IR{\relax{\rm I\kern-.18em R}}

\lref\Mal{
  J.~M.~Maldacena,
  ``The large N limit of superconformal field theories and supergravity,''
  Adv.\ Theor.\ Math.\ Phys.\  {\bf 2}, 231 (1998)
  [Int.\ J.\ Theor.\ Phys.\  {\bf 38}, 1113 (1999)]
  [arXiv:hep-th/9711200].
}

\lref\Nielsen{ See for example, M.\ Nielsen, and I.\ L.\ Chuang,
``Quantum Computation and Quantum Information'', Cambridge
university press, (2000).}

\lref\GO{D.~J.~Gross and H.~Ooguri,
  ``Aspects of large N gauge theory dynamics as seen by string theory,''
  Phys.\ Rev.\ D {\bf 58}, 106002 (1998)
  [arXiv:hep-th/9805129].
}

\lref\DGO{N.~Drukker, D.~J.~Gross and H.~Ooguri,
  ``Wilson loops and minimal surfaces,''
  Phys.\ Rev.\ D {\bf 60}, 125006 (1999)
  [arXiv:hep-th/9904191].
}

\lref\casinifree{
 H.~Casini, C.~D.~Fosco and M.~Huerta,
   ``Entanglement and alpha entropies for a massive Dirac field in two
  dimensions,''
  J.\ Stat.\ Mech.\  {\bf 0507}, P007 (2005)
  [arXiv:cond-mat/0505563];
H.~Casini and M.~Huerta,
   ``Entanglement and alpha entropies for a massive scalar field in two
  dimensions,''
  J.\ Stat.\ Mech.\  {\bf 0512}, P012 (2005)
  [arXiv:cond-mat/0511014].
}

\lref\Casinifour{H.~Casini and M.~Huerta,
  ``A finite entanglement entropy and the c-theorem,''
  Phys.\ Lett.\ B {\bf 600}, 142 (2004)
  [arXiv:hep-th/0405111].
}

\lref\Casinithree{
 H.~Casini,
  ``Geometric entropy, area, and strong subadditivity,''
  Class.\ Quant.\ Grav.\  {\bf 21}, 2351 (2004)
  [arXiv:hep-th/0312238].
}

\lref\CHcusp{
  H.~Casini and M.~Huerta,
  ``Universal terms for the entanglement entropy in 2+1 dimensions,''
  arXiv:hep-th/0606256.
}

\lref\Furold{
  D.~V.~Fursaev,
   ``Entanglement entropy in critical phenomena and analogue models of quantum
  gravity,''
  Phys.\ Rev.\ D {\bf 73}, 124025 (2006)
  [arXiv:hep-th/0602134].
}

\lref\Em{
  R.~Emparan,
  ``Black hole entropy as entanglement entropy: A holographic derivation,''
  JHEP {\bf 0606}, 012 (2006)
  [arXiv:hep-th/0603081].
}

\lref\Iwa{
  Y.~Iwashita, T.~Kobayashi, T.~Shiromizu and H.~Yoshino,
  ``Holographic entanglement entropy of de Sitter braneworld,''
  arXiv:hep-th/0606027.
}

\lref\RT{
  S.~Ryu and T.~Takayanagi,
  ``Holographic derivation of entanglement entropy from AdS/CFT,''
  Phys.\ Rev.\ Lett.\  {\bf 96}, 181602 (2006)
  [arXiv:hep-th/0603001].
}

\lref\RTL{
  S.~Ryu and T.~Takayanagi,
  ``Aspects of holographic entanglement entropy,''
  arXiv:hep-th/0605073.
}

\lref\Fur{
  D.~V.~Fursaev,
  ``Proof of the holographic formula for entanglement entropy,''
  arXiv:hep-th/0606184.
}

\lref\Sol{
  S.~N.~Solodukhin,
  ``Entanglement entropy of black holes and AdS/CFT correspondence,''
  arXiv:hep-th/0606205.
}

\lref\Bom{L.~Bombelli, R.~K.~Koul, J.~H.~Lee and R.~D.~Sorkin,
  ``A Quantum Source of Entropy for Black Holes,''
  Phys.\ Rev.\ D {\bf 34}, 373 (1986);
M.~Srednicki,
  ``Entropy and area,''
  Phys.\ Rev.\ Lett.\  {\bf 71}, 666 (1993)
  [arXiv:hep-th/9303048].
}

\lref\GKP{ S.~S.~Gubser, I.~R.~Klebanov and A.~M.~Polyakov,
  ``Gauge theory correlators from non-critical string theory,''
  Phys.\ Lett.\ B {\bf 428}, 105 (1998)
  [arXiv:hep-th/9802109];

E.~Witten,
  ``Anti-de Sitter space and holography,''
  Adv.\ Theor.\ Math.\ Phys.\  {\bf 2}, 253 (1998)
  [arXiv:hep-th/9802150].
}

\lref\Bousso{
  R.~Bousso,
  ``A Covariant Entropy Conjecture,''
  JHEP {\bf 9907}, 004 (1999)
  [arXiv:hep-th/9905177];
  ``Holography in general space-times,''
  JHEP {\bf 9906}, 028 (1999)
  [arXiv:hep-th/9906022].
}

\lref\susskind{
 L.~Susskind,
  ``The World As A Hologram,''
  J.\ Math.\ Phys.\  {\bf 36}, 6377 (1995)
  [arXiv:hep-th/9409089].
}

\lref\Cardy{C.~Holzhey, F.~Larsen and F.~Wilczek,
  ``Geometric and renormalized entropy in conformal field theory,''
  Nucl.\ Phys.\ B {\bf 424}, 443 (1994)
  [arXiv:hep-th/9403108];

P.~Calabrese and J.~L.~Cardy,
  ``Entanglement entropy and quantum field theory,''
  J.\ Stat.\ Mech.\  {\bf 0406}, P002 (2004)
  [arXiv:hep-th/0405152].
}

\lref\Za{K.~Zarembo,
  ``Wilson loop correlator in the AdS/CFT correspondence,''
  Phys.\ Lett.\ B {\bf 459}, 527 (1999)
  [arXiv:hep-th/9904149];
P.~Olesen and K.~Zarembo, ``Phase transition in Wilson loop
correlator from AdS/CFT correspondence,'' arXiv:hep-th/0009210.
}

\lref\Kim{
 H.~Kim, D.~K.~Park, S.~Tamarian and H.~J.~W.~Muller-Kirsten,
   ``Gross-Ooguri phase transition at zero and finite temperature: Two  circular
  Wilson loop case,''
  JHEP {\bf 0103}, 003 (2001)
  [arXiv:hep-th/0101235].
}

\lref\Wilson{
  J.~M.~Maldacena,
  ``Wilson loops in large N field theories,''
  Phys.\ Rev.\ Lett.\  {\bf 80}, 4859 (1998)
  [arXiv:hep-th/9803002];
  S.~J.~Rey and J.~T.~Yee,
   ``Macroscopic strings as heavy quarks in large N gauge theory and  anti-de
  Sitter supergravity,''
  Eur.\ Phys.\ J.\ C {\bf 22}, 379 (2001)
  [arXiv:hep-th/9803001].
}

\lref\Moore{
  E.~Fradkin and J.~E.~Moore,
   ``Entanglement entropy of 2D conformal quantum critical points: hearing the
  shape of a quantum drum,''
  arXiv:cond-mat/0605683.
}

\lref\Solc{
  S.~N.~Solodukhin,
  ``The Conical singularity and quantum corrections to entropy of black hole,''
  Phys.\ Rev.\ D {\bf 51}, 609 (1995)
  [arXiv:hep-th/9407001];
  ``On 'Nongeometric' contribution to the entropy of black hole due to quantum
  corrections,''
  Phys.\ Rev.\ D {\bf 51}, 618 (1995)
  [arXiv:hep-th/9408068];
  ``One loop renormalization of black hole entropy due to nonminimally coupled
  matter,''
  Phys.\ Rev.\ D {\bf 52}, 7046 (1995)
  [arXiv:hep-th/9504022].
}

\lref\AL{
H.~Araki and E.~H.~Lieb,
  ``Entropy inequalities,''
  Commun.\ Math.\ Phys.\  {\bf 18}, 160 (1970).
}

\lref\strong{
  E.~H.~Lieb and M.~B.~Ruskai,
  ``Proof of the strong subadditivity of quantum-mechanical entropy,''
  J.\ Math.\ Phys.\  {\bf 14}, 1938 (1973).
}

\lref\KS{ I.~R.~Klebanov and A.~A.~Tseytlin,
  ``Gravity duals of supersymmetric SU(N) x SU(N+M) gauge theories,''
  Nucl.\ Phys.\ B {\bf 578}, 123 (2000)
  [arXiv:hep-th/0002159];
I.~R.~Klebanov and M.~J.~Strassler,
   ``Supergravity and a confining gauge theory: Duality cascades and
  chiSB-resolution of naked singularities,''
  JHEP {\bf 0008}, 052 (2000)
  [arXiv:hep-th/0007191].
}

\lref\MaE{
  J.~M.~Maldacena,
  ``Eternal black holes in Anti-de-Sitter,''
  JHEP {\bf 0304}, 021 (2003)
  [arXiv:hep-th/0106112].
}

\lref\HMS{
  S.~Hawking, J.~M.~Maldacena and A.~Strominger,
  ``DeSitter entropy, quantum entanglement and AdS/CFT,''
  JHEP {\bf 0105}, 001 (2001)
  [arXiv:hep-th/0002145].
}

\lref\AFN{J.~Aczel, B.~Forte and C.~T.~Ng, ``Why the Shannon and
Hartley entropies are natural,'' Adv. Appl. Prob. {bf 6} (1974),
131;

W.~Ochs, ``A new axiomatic characterization of the von Neumann
entropy,'' Rep. Math. Phys. {\bf 8} (1975), 109;

A. Wehrl, ``General properties of entropy,'' Rev. Mod. Phys. {\bf
50}, 1978, 221.}

\lref\Ba{
  C.~Bachas,
  ``Convexity Of The Quarkonium Potential,''
  Phys.\ Rev.\ D {\bf 33}, 2723 (1986).
}

\lref\GOW{
  J.~Greensite and P.~Olesen,
   ``Worldsheet fluctuations and the heavy quark potential in the AdS/CFT
  approach,''
  JHEP {\bf 9904}, 001 (1999)
  [arXiv:hep-th/9901057].
}

\lref\DP{
  H.~Dorn and V.~D.~Pershin,
   ``Concavity of the Q anti-Q potential in N = 4 super Yang-Mills gauge  theory
  and AdS/CFT duality,''
  Phys.\ Lett.\ B {\bf 461}, 338 (1999)
  [arXiv:hep-th/9906073].
}

\lref\DF{
  N.~Drukker and B.~Fiol,
   ``On the integrability of Wilson loops in AdS(5) x S**5: Some periodic
  ansatze,''
  JHEP {\bf 0601}, 056 (2006)
  [arXiv:hep-th/0506058].
}

\baselineskip 18pt plus 2pt minus 2pt

\Title{\vbox{\baselineskip12pt \hbox{hep-th/0608213}
\hbox{KUNS-2038}
  }}
{\vbox{\centerline{AdS/CFT and Strong Subadditivity of Entanglement
Entropy} }}

\centerline{Tomoyoshi
Hirata\foot{e-mail:hirata@gauge.scphys.kyoto-u.ac.jp} and Tadashi
Takayanagi\foot{e-mail:takayana@gauge.scphys.kyoto-u.ac.jp}}
\medskip\centerline{Department of Physics, Kyoto University, Kyoto
606-8502, Japan}

\vskip .5in \centerline{\bf Abstract} Recently, a holographic
computation of the entanglement entropy in conformal field theories
has been proposed via the AdS/CFT correspondence. One of the most
important properties of the entanglement entropy is known as the
strong subadditivity. This requires that the entanglement entropy
should be a concave function with respect to geometric parameters.
It is a non-trivial check on the proposal to see if this property is
indeed satisfied by the entropy computed holographically. In this
paper we examine several examples which are defined by annuli or
cusps, and confirm the strong subadditivity via direct calculations.
Furthermore, we conjecture that Wilson loop correlators in strongly
coupled gauge theories satisfy the same relation. We also discuss
the relation between the holographic entanglement entropy and the
Bousso bound.

\noblackbox

\Date{August, 2006}

\writetoc

\newsec{Introduction}

In order to understand quantum gravity, we need to know the
structure of its Hilbert space ${\cal{H}}_{tot}$ and the quantum
state $|\Psi\lb$ which describes a given background. In this setup
one of the most basic quantities is considered to be the
entanglement entropy. Let us divide the total Hilbert space into a
direct product of two subspaces ${\cal{H}}_{tot}={\cal{H}}_A\otimes
{\cal{H}}_B$. The entanglement entropy is defined as the von Neumann
entropy when we trace out the subspace ${\cal{H}}_B$. This measures
the information loss accompanied with this smearing process. The
entropy depends only on the total density matrix
$\rho_{tot}=|\Psi\lb\la\Psi|$ and the structure of the Hilbert
space. Therefore it is a universal and basic quantity to
characterize the quantum state as we do not have to make explicit
the details of the theory such as matter contents and so on.

The modern understanding of holography, especially in the AdS/CFT
correspondence \Mal ,  tells us that the Hilbert space
${\cal{H}}_{tot}$ of the gravity theory is actually equivalent to
the Hilbert space of a certain quantum field theory (QFT) that lives
on the boundary of the spacetime. In AdS/CFT, the QFT becomes a
conformal field theory (CFT) and this is the setup we consider in
this paper mainly.

In QFTs, we have two Hilbert subspaces by dividing the total space
manifold into two submanifolds $A$ and $B$. The entanglement entropy
is defined geometrically by tracing out the states which live on the
submanifold $A$ \Bom. The holography argues that the Hilbert space
of its dual gravity theory can also be written as a direct product
as is necessary to define the entanglement entropy. Moreover, we
expect that the entanglement entropy computed in the dual field
theory should be the same as the one in the gravity theory.

Recently, a holographic computation of the entanglement entropy is
proposed in \RT \RTL, based on the AdS/CFT correspondence \Mal.
There, the entropy is calculated by replacing the horizon area in
the Bekenstein-Hawking formula with the area of minimal surface in
AdS space whose boundary is the same as that of the submanifold $A$.
This offers us a powerful way to calculate the entanglement entropy;
otherwise it involves the complicated quantum analysis, which is
very hard except two dimensional CFTs \Cardy. This relation has been
successfully applied to black holes in brane-world  \Em \Sol\ and
de-Sitter spaces \Iwa\ (refer to \HMS\ and \MaE\ for earlier
discussions). A slightly heuristic proof of this proposal has been
given in \Fur\ based on the first principle of AdS/CFT
correspondence \GKP. Moreover, as we will point out in this paper,
the above holographic computation of entanglement entropy is closely
related to the covariant entropy bound known as the Bousso bound
\Bousso. We will argue that the holographic computation saturates
this bound. This consideration makes clear the following basic
question in AdS/CFT correspondence: {\it what part of the AdS space
encodes the information of the boundary CFT included in a specific
region}?

In this paper we would like to explore this proposal and to find
further evidences from a different viewpoint. One of the most
important properties which are satisfied by the entanglement entropy
is called the strong subadditivity \strong. This represents a sort
of irreversibility of entropy and is considered to be the strongest
condition of entanglement entropy. Indeed, it has rigorously been
shown that the strong subadditivity with several other more obvious
conditions characterizes the von-Neumann entropy. Moreover, it is
known that the entropic analogue of Zamolodchikov's $c-$theorem can
be derived from the strong subadditivity in 2D QFTs \Casinifour
\Casinithree.

The main purpose of this paper is to check if this condition is
satisfied in the holographic calculations in several explicit
examples. In particular we consider the case where the submanifold
$A$ is given by an annulus, and also the case where its boundary
$\de A$ is a line with a cusp singularity. The latter examples in
the (2+1) D free scalar field theory were discussed in \CHcusp\
quite recently (refer to \Furold\ for a general discussion and also
to \Moore\ for an analysis at quantum critical point with a
conformal invariant ground state
       wavefunction). Our analysis corresponds to the result
       in the strongly coupled
supersymmetric gauge theories of the same dimension. The strong
subadditivity requires that a specific term in the entropy should be
a concave function\foot{ As usual, we call a function $f$ is concave
(or convex) if $f''\leq 0$ (or $f''\geq 0$). Even though the
specific terms in the entropy itself are concave, we will also
encounter convex functions because we sometimes flip the sign of the
terms.} when we change the value of the geometrical quantity such as
the ratio of the radius in the annulus and the deficit angle of a
cusp. Indeed, in all our examples we will find that the strong
subadditivity is satisfied. This provides an additional evidence of
the proposed holographic derivation of the entanglement entropy. We
can also regard it as an evidence of the AdS/CFT correspondence
itself.

At the same time, as a bonus, we will notice the similarity between
the holographic calculation of the entanglement entropy \RT\ and the
well-known evaluation of the Wilson loops expectation values in
AdS/CFT \Wilson. In particular, we can easily find that the
entanglement entropy in a $(2+1)$ dimensional CFT is equivalent to
the Wilson loop in a $(3+1)$ dimensional CFT defined by a strongly
coupled gauge theory. Thus we conjecture that the strong
subadditivity relation is also true for strongly coupled gauge
theories.

This paper is organized as follows. In section two, we review the
basic definition and properties of the entanglement entropy such as
the strong subadditivity. In section three, we apply the strong
subadditivity to various examples of the entanglement entropy in
CFTs. In section four, we explain the holographic computation of the
entanglement entropy via the AdS/CFT. We also relate it to the
covariant entropy bound (so called the Bousso bound). In section
five, we examine specific examples of the holographic entanglement
entropy and check if it satisfies the strong subadditivity. In
section six, we summarize conclusions.

\newsec{Entanglement Entropy and Strong Subadditivity}

\subsec{Definition of Entanglement Entropy}

Consider a quantum mechanical system with many degrees of freedom
such as spin chains. More generally, we can consider arbitrary
lattice models or quantum field theories\foot{The entanglement
entropy is generally divergent in continuum theories. Therefore
usually we assume an ultraviolet cutoff $a$ to regulate the quantum
field theory. In this sense, strictly speaking, the entropy should
always be defined in the regularized lattice version of a given
quantum field theories. Below we assume that this is just a
technical issue and that we can always have such a regularization.}
(QFTs) including conformal field theories (CFTs). We put the system
at zero temperature and the total quantum system is described by the
pure ground state $|\Psi\lb$. We assume no degeneracy of the ground
state. Then, the density matrix is that of the pure state \eqn\pure{
\rho_{tot}=|\Psi\rangle \langle \Psi|.} The von Neumann entropy of
the total system is clearly zero $ S_{tot}= -\tr\, \rho_{tot} \log
\rho_{tot}=0$.

Next we divide the total system into two subsystems $A$ and $B$. In
the spin chain example, we artificially cut off the chain at some
point and divide the lattice points into two groups. Notice that
physically we do not do anything to the system and the cutting
procedure is an imaginary process. Accordingly the total Hilbert
space can be written as a direct product of two spaces
${\cal{H}}_{tot}={\cal{H}}_{A}\otimes {\cal{H}}_{B}$ corresponding
to those of subsystems $A$ and $B$. Let ${\cal O}_A$ be an operator
which acts non-trivially only on $A$. Then its expectation value is
\eqn\expec{\la {\cal O}_A \lb= \tr {\cal O}_A\cdot
\rho_{tot}=\tr_{A}{\cal O}_A\cdot \rho_{A},} where the trace $\tr_A$
is taken only over the Hilbert space ${\cal{H}}_{A}$. Here we
defined the reduced density matrix $\rho_A$ defined by \eqn\ra{
\rho_A= \tr_{B}~\rho_{tot}, } by tracing out the Hilbert space
${\cal{H}}_{B}$. Thus the observer who is only accessible to the
subsystem $A$ feels as if the total system were described by the
reduced density matrix $\rho_A$.

Then we define the entanglement entropy of the subsystem $A$ as the
von Neumann entropy of the reduced density matrix $\rho_A$
\eqn\denenatangle{S_A= - \tr_{A}\, \rho_{A} \log \rho_{A}.} This
entropy measures the amount of information lost by tracing out (or
smearing) the subsystem $B$.

It is also possible to define the entanglement entropy $S_A(\beta)$
at finite temperature $T=\beta^{-1}$. This can be done just by
replacing \pure\ with the thermal one $\rho_{thermal}=e^{-\beta H}$,
where $H$ is the total Hamiltonian. When $A$ is the total system,
$S_A(\beta)$ is clearly the same as the thermal entropy.

\subsec{Strong Subadditivity}

The entanglement entropy enjoys several useful properties. When the
system is at zero temperature (i.e. pure state), it is easy to show
\eqn\intensive{S_A=S_B,} which manifestly shows that the entropy is
not extensive as opposed to the ordinary thermal entropy.
              This equality \intensive\ is violated
              for a mixed state (e.g. finite temperature).

Below we assume that the state is not pure in general. When we start
with two subsystems $A$ and $A'$, we can show
\eqn\subadd{S_{A}+S_{A'}\geq S_{A\cup A'}.} This is called
subadditivity. We find this relation intuitively clear since the
information for the overlapped part $A\cap A'$ is double counted in
the left-hand side of \subadd. It is true even when $A\cap A'$ is
empty.

Actually, a stronger inequality is known to be satisfied (again we
assume the mixed state generally).
 This is called the strong subadditivity
\strong\ given by the following formula \eqn\strongg{S_{A} + S_{A'}
\ge S_{A\cup A'}+S_{A\cap A'}.} Equally we can write this inequality
as follows. Divide the subsystem $A$ into three parts $A_1,A_2$ and
$A_3$ such that each of them does not intersect with each other,
i.e. \eqn\partthree{{\cal H}_A={\cal H}_{A_1}\otimes {\cal
H}_{A_2}\otimes {\cal H}_{A_3},} then we can show
\eqn\strongtwo{S_{A_1+A_2+A_3} + S_{A_2} \le
S_{A_1+A_2}+S_{A_2+A_3}.} Notice that if there is no correlation
between the three Hilbert spaces (i.e. $\rho_{tot}=\rho_{A_1}\otimes
\rho_{A_2}\otimes \rho_{A_3}$), the both sides in \strongtwo\ become
equal.

By multiplying an extra Hilbert space $A_4$ with $A_1$, $A_2$ and
$A_3$ such that the system is pure with respect to the total Hilbert
space ${\cal H}_{A_{tot}}={\cal H}_{A_1}\otimes {\cal
H}_{A_2}\otimes {\cal H}_{A_3}\otimes {\cal H}_{A_4}$, we can find
another inequality equivalent to \strongtwo\ \eqn\strongsban{
S_{A_1}+S_{A_3}\leq S_{A_1+A_2}+S_{A_2+A_3}.} It is easy to see that
the subadditivity \subadd\ and the triangle inequality (Araki-Lieb
inequality) \AL \eqn\suble{|S_{A_1}-S_{A_2}|\leq S_{A_1+ A_2},} can
be derived from (i.e. weaker than) \strongtwo\ and \strongsban.

The strong subadditivity \strongg \strongtwo \strongsban\ is also
satisfied for the classical entropy known as the Shannon entropy and
in this case the proof is almost the same as the proof of the
classical version of the subadditivity \subadd . However, the proof
of the strong subadditivity for the quantum entropy (von Neumann
entropy) is rather complicated. Indeed the original proof given by
Lieb and Ruskai \strong\ is highly non-trivial.

To see an outlook of the proof for the strong subadditivity we begin
with a definition of jointly concavity (we will follow the proof in
\Nielsen). Suppose $f(A,B)$ is a real-valued function of two
matrices $A$ and $B$. The function $f$ is said to be jointly concave
in $A$ and $B$ if \eqn\jointly{f(\lambda A_1 +(1-\lambda)A_2,\lambda
B_1 +(1-\lambda )B_2)\ge \lambda f(A_1,B_1)+(1- \lambda
)f(A_2,B_2),} for all $0\le \lambda \le 1$.

To prove the strong subadditivity, we firstly use the Lieb's
theorem. Lieb's theorem states \eqn\lieb{f(A,B) \equiv \tr
(X^\dagger A^t X B^{1-t} )} is jointly concave for any matrix $X$
and for all $0 \le t \le 1$ in positive matrices $A$ and $B$. Then
Lieb's theorem implies the joint convexity of the relative entropy
\eqn\relative{S(\rho||\sigma) \equiv \tr (\rho \log \rho)- \tr (\rho
\log \sigma).} in $\sigma$ and $\rho$, and the joint convexity of
the relative entropy shows that the conditional entropy
\eqn\conditonal{S(A|B)\equiv S_{A+B} -S_B} is concave in
$\rho_{A+B}$. Finally, strong subadditivity \strongg \strongtwo
\strongsban\ follows from the concavity of the conditional entropy.
More details of the proof and other properties of the entanglement
entropy can be found in e.g. \Nielsen.

 As is clear from the above proof, the
strong subadditivity has a deep relationship with the concavity of
entropy. It will become clear later that the strong subadditivity
tells us the entropy is concave when we change the values of
parameters that determine the subsystems. It is the strongest
condition ever found for the von-Neumann entropy. Indeed, it has
mathematically been shown that the strong subadditivity with several
other more obvious conditions (such as the invariance under unitary
transformations and the continuity with respect to the eigenvalues
of $\rho_{tot}$) characterizes the von-Neumann entropy \AFN.

\subsec{Entanglement Entropy in Relativistic Theories}

As we have explained, the entanglement entropy is defined by
dividing the Hilbert space into two subspaces. In a $(d+1)$
dimensional quantum field theory, this can be done geometrically by
specifying a $d$ dimensional submanifold $B$ embedded in the total
$d$ dimensional space. Thus below we use the same symbols $A,B,...$
to specify both the subsystem and the submanifold at the same time.
 We assume that the spacetime is simply given by the
flat space $R^{1,d}$ just for simplicity.

In the relativistic theory we have to take care of the Lorentz
invariance. We can take the space-like surface at fixed time $t=t_0$
as a Cauchy surface and divide it into $A$ and $B$. This manifestly
divides the total Hilbert space ${\cal H}_{tot}$ into two subspaces
${\cal H}_A$ and ${\cal H}_B$ such that ${\cal H}={\cal H}_A\otimes
{\cal H}_B$. The entanglement entropy $S_A$ is define by taking the
trace over states living on the submanifold $B$.

However, we can take other Cauchy surfaces to realize the same
partition of the Hilbert space. Consider the surface $B'$ instead of
$B$ in Fig.1. This leads to the same Hilbert subspace i.e. ${\cal
H}_{B'}={\cal H}_B$ because the physics on $B'$ is completely
determined if we fix the initial condition on $B$. The information
on $B$ determines $D^+(B)$, which is called the future domain of
dependence. The boundary of $D^+(B)$ is the (future) Cauchy horizon
$H^+(B)$. Therefore, the Hilbert subspace ${\cal H}_B$ is specified
if we choose the Cauchy horizon $H^+(B)$ \Casinithree.

\fig{The subspaces $B$ and $B'$ of the Cauchy surfaces define the
same Hilbert space ${\cal H}_{B'}={\cal H}_B$. This consideration
leads to the conclusion that the Hilbert space is classified by the
Cauchy horizon $H^+(B)$.}{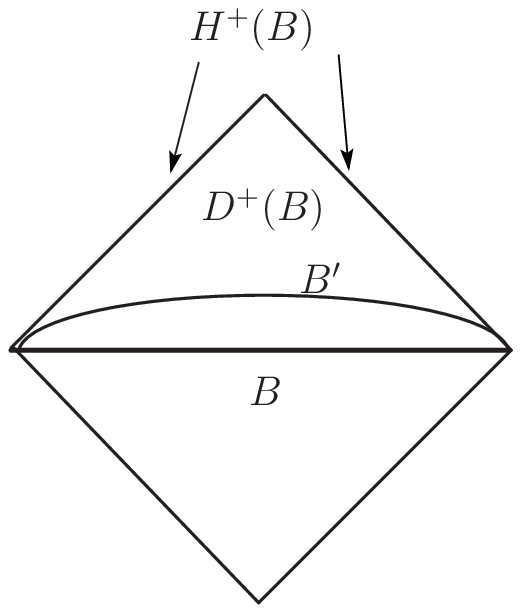}{1.5truein}

\newsec{Applications of Strong Subadditivity}

Here we would like to apply the strong subadditivity relation
\strongg \strongtwo\  to the entanglement entropy $S_A$ in conformal
field theories in order to find constraints on the properties of
$S_A$. Especially we are interested in $(2+1)$ and $(3+1)$
dimensional CFTs, though it is not difficult to extend our argument
to higher dimensional theories.

\subsec{Entanglement Entropy in the Presence of Cusp}

Consider the entanglement entropy for a $(2+1)$ dimensional CFT. We
assume that the spacetime is simply given by $R^{1,2}$. The
subsystem $A$ (at a fixed time $t=0$) is geometrically described by
a submanifold $A$ in the two dimensional space-like manifold $R^2$.
When the boundary $\de A$ is smooth (e.g. the case where $\de A$ is
a circle), we find the following behavior of the entropy
\eqn\onetwodim{S_A=\gamma\cdot \f{|\de A|}{a} +b,} where $a$ is the
UV cutoff (lattice spacing), and $|\de A|$ denotes the length of the
boundary of $A$. Also $\gamma$ is a numerical constant. In general
the leading divergent term is proportional to the area of the
boundary $\de A$, which is known as the area law \Bom. The first
term in \onetwodim\ corresponds to this contribution. The constant
$b$ does not depend on the cutoff $a$ and thus can be considered as
a universal quantity. It depends on the explicit form of $\de A$ in
a conformal invariant way.

On the other hand, if the boundary $\de A$ is singular and has a
cusp,
 there exists a logarithmic correction as found
quite recently in various systems \CHcusp \Moore. The cusp is
specified by an angle defined such that $\Omega=\pi$ corresponds to
a smooth line (see Fig.2). Therefore in this case the entropy takes
the form \eqn\onetwologdim{S_A=\gamma\cdot \f{|\de
A|}{a}+f(\Omega)\log a+ b.} Notice that the constant term $b$
depends on the cutoff $a$ in this case.

\fig{The boundary with a cups (left). The setup to which the strong
subadditivity is applied (right).}{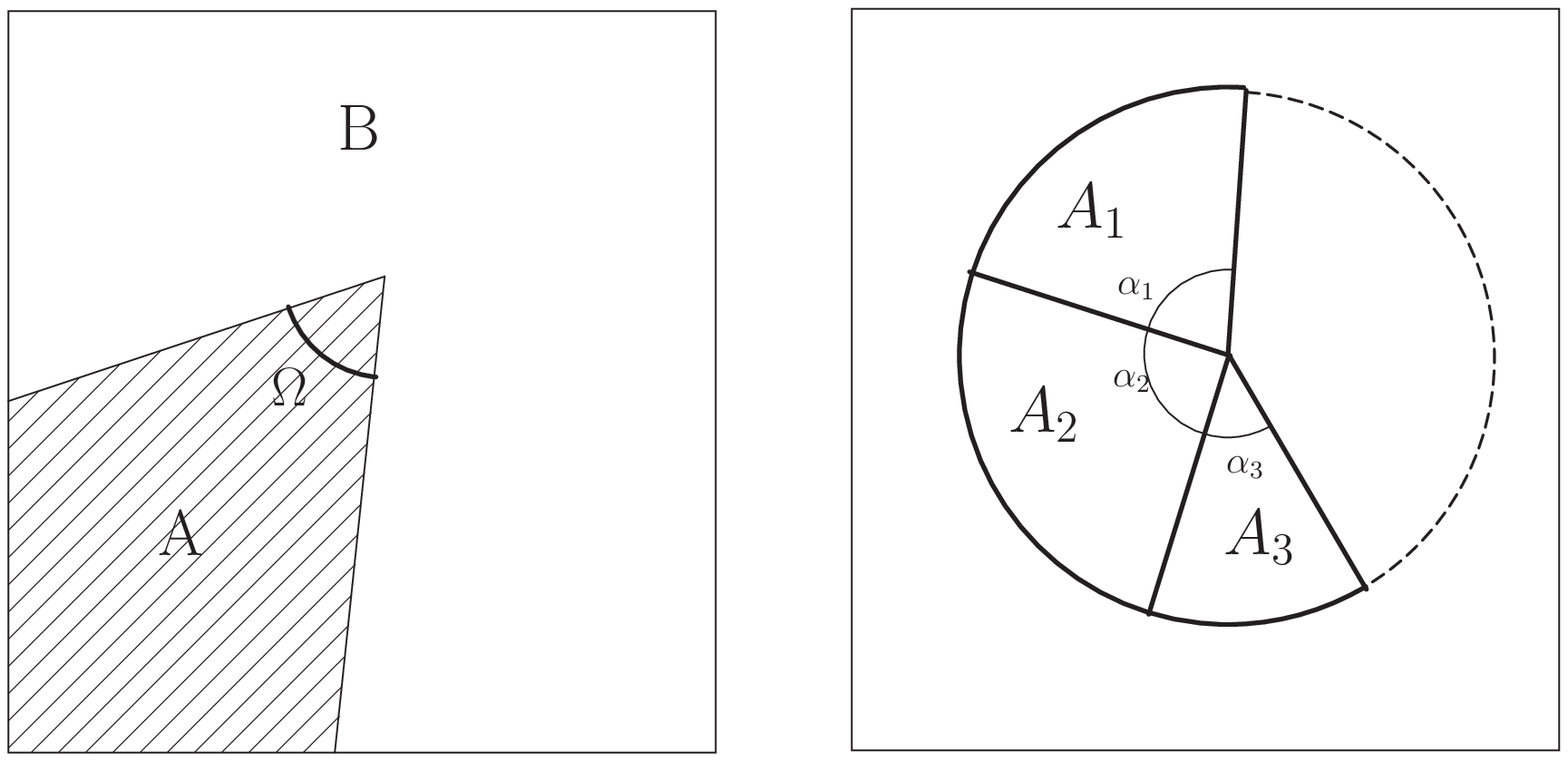}{4truein}

Now we would like to derive constraints on $f(\Omega)$ from the
basic properties of entanglement entropy. First of all, by
definition we find \eqn\defggh{f(\pi)=0.} Also the equality
$S_A=S_B$ \pure\ leads to \eqn\reflection{f(\Omega)=f(2\pi-\Omega).}

Next we would like to apply the strong subadditivity. Assume that
the submanifolds $A_1$, $A_2$ and $A_3$ each have cusps with the
angles $\ap_1,\ap_2$ and $\ap_3$ such that one side of the cusp of
$A_1$ (or $A_2$) is glued with that of $A_2$ (or $A_3$),
respectively (thus we require $\sum_i\ap_i<2\pi$) as in Fig.2. Then
the strong subadditivity leads to (note that leading divergences are
all canceled out and that we finally take the limit $a\to 0$)
\eqn\strongcusp{f(\ap_1+\ap_2+\ap_3)-f(\ap_1+\ap_2)\geq
f(\ap_2+\ap_3) -f(\ap_2).} By taking the limit $\ap_3\to +0$ we
obtain \eqn\strongder{f'(\ap_1+\ap_2)\geq f'(\ap_2).} Finally the
limit $\ap_1\to +0$ leads to \eqn\strongderr{f''(\Omega)\geq 0.}
Thus $f(\Omega)$ is a convex function\foot{Even though the entropy
is concave, the function $f$ is convex since the pre-factor $\log a$
is negative.}. Furthermore, the opposite is true, i.e. if $f$ is
convex, then \strongcusp\ does hold.

It is also possible to combine the relation \reflection\ with the
strong subadditivity. Especially we take $\ap_1=2\pi-2\ap_2$. Then
\strongcusp\ can be reduced to $f(\ap_2-\ap_3) \geq f(\ap_2+\ap_3)$.
Thus we find \eqn\condlston{f'(\Omega)\leq 0 \ \ \ (\rm{if}\ \
\Omega\leq \pi).} This leads to $f(\Omega)\geq 0$ due to
\defggh.

In conclusion, our results \defggh, \strongderr\ and \condlston\ can
be summarized as follows \eqn\finalstrong{f(\Omega)\geq 0, \ \ \
f(\pi)=0,\ \ \ f'(\Omega)\leq 0, \ \ \  f''(\Omega)\geq 0,\ \ \
(0\leq \Omega \leq \pi).} For the values $\pi\leq \Omega \leq 2\pi$
we can use the identity $f(\Omega)=f(2\pi-\Omega)$.

\subsec{Entanglement Entropy for Annular Subsystem}

Next we turn to the case where the submanifold $A$ is an annulus
 such that its boundary $\de A$ consists of two concentric rings
 with the radius $r_1$ and $r_2$ (assume $r_1<r_2$) as in Fig.3. We define its
 entanglement entropy by $S_{A}(r_1,r_2)$. As is clear from the
 discussions in the previous subsection, it takes the following
 general form \eqn\entrosa{S_A(r_1,r_2)=\gamma\cdot
 \f{2\pi(r_1+r_2)}{a}+b\left(\f{r_2}{r_1}\right).} Notice that the universal finite
 term $b$ is a function of $r_2/r_1$ since it is dimensionless.
 There is no logarithmic term as opposed to the previous example
 because the boundary does not include any cusp singularity.

\fig{The annulus boundary (left). The setup to which the strong
subadditivity is applied (right).}{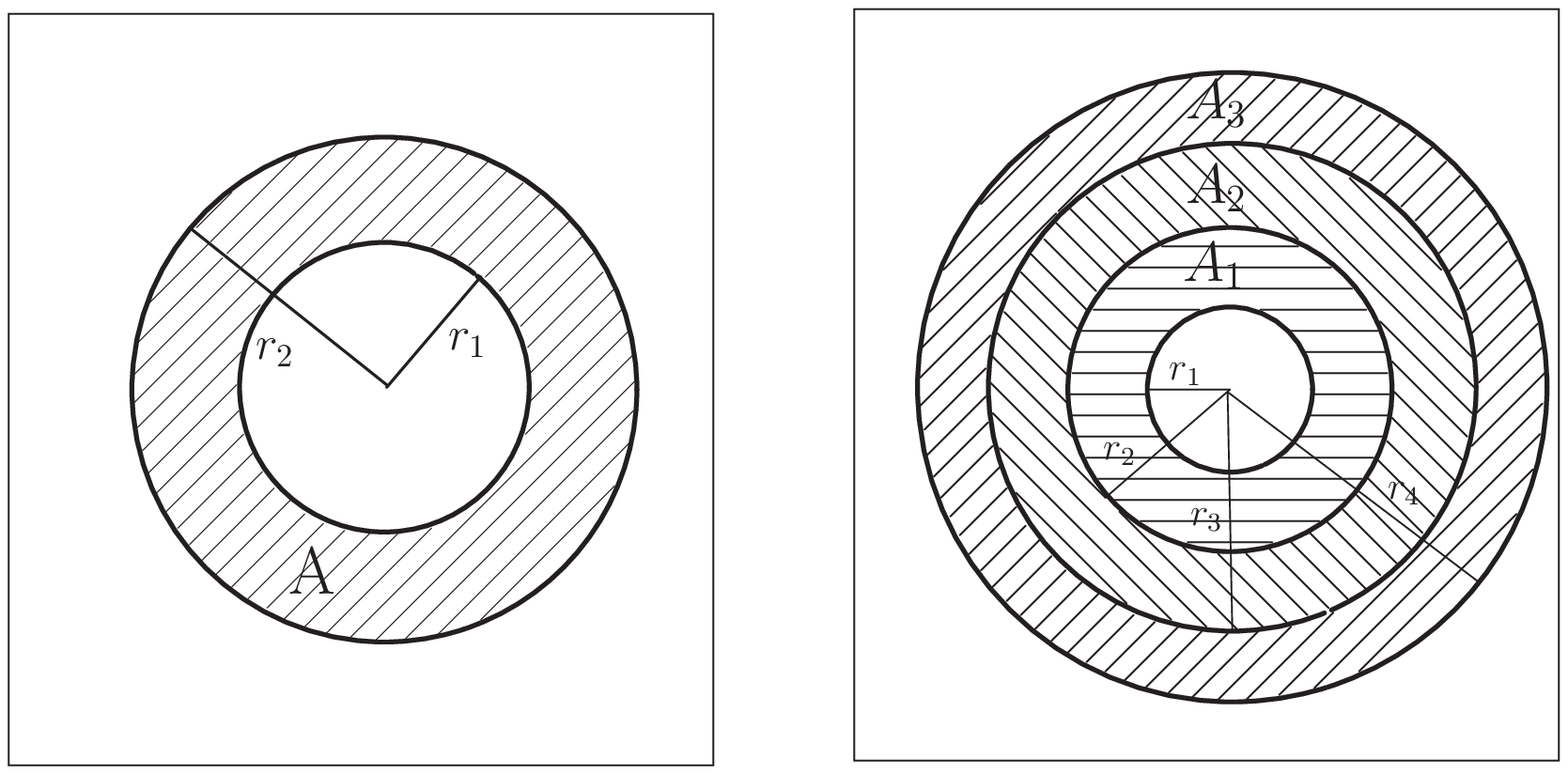}{4truein}

Now we would like to apply the strong subadditivity. Suppose three
annular submanifolds $A_1$, $A_2$ and $A_3$ are placed successively
such that the radii of their concentric rings are $(r_1,r_2)$,
$(r_2,r_3)$ and $(r_3,r_4)$, respectively ($r_1<r_2<r_3<r_4$) as in
Fig.3.

It is also useful to define $\rho_i=\log r_i$. Then it is
straightforward to write down the inequality obtained from the
strong subadditivity \eqn\stronb{b(\rho_4-\rho_1)+b(\rho_3-\rho_2)
\leq b(\rho_3-\rho_1)+b(\rho_4-\rho_2).}

As a particular limit $\rho_4=\infty$ we find $b(\rho_3-\rho_2) \leq
b(\rho_3-\rho_1).$ Thus we find \eqn\deristr{ b'(\rho)\geq 0.} Also
by taking the limits $\rho_4\to \rho_3$ and $\rho_2\to \rho_1$, we
find \eqn\strondertwp{b''(\rho)\leq 0.} Furthermore, the limit
$\rho\to \infty$ is equivalent to the case where $A$ is a circular
disk since the radius $r$ of the inner hole goes to zero. We write
its value of $b$ as $b_{disk}$. In summary we have found the
following properties of the function $b(\rho)$ ($\rho=\log
\f{r'}{r}>0$) \eqn\propertyb{b(\rho)\leq b(\infty)=b_{disk},\ \ \
b'(\rho)\geq 0, \ \ \ b''(\rho)\leq 0.}

It is also intriguing to examine higher dimensional cases. The
analogous boundary $\de A$ can be obtained by replacing the two
concentric rings with two such $d-1$ dimensional spheres. In
particular, we concentrate on the most interesting case i.e. $d=3$
(4D CFT). The corresponding entanglement entropy takes the following
form \eqn\entansth{S_A=\gamma\cdot \f{r_1^2+r_2^2}{a^2}
-f\cdot\log\left(\f{r_1r_2}{a^2}\right)+b\left(\f{r_2}{r_1}\right).}
The coefficient $f$ in front of the logarithmic term is universal
and it is proportional to a linear combination of central charges in
the 4D CFT \RTL\ (see also \Solc\ for earlier discussions). The
general formula derived in \RTL\ and the dual holographic
computation via AdS/CFT discussed later show that $f=f(r_2/r_1)$
does not actually depend on $r_1/r_2$, i.e. just a constant. This is
also clear from the observation that this term should be a local
quantity since it originally depends on the cutoff $a$ and is
divergent. Even though the function $b$ depends on the choice of the
cutoff $a$, we can apply the strong subadditivity as before and
obtain the properties \propertyb\ with $b(\infty)=b_{disk}$ now
replaced by $b(\infty)=b_{sphere}$. This is because both quadratic
and logarithmic divergent term depend on the radii $r_1$ and $r_2$
only in the form $g(r_1)+g(r_2)$ and thus they are completely
canceled out in the inequality of the strong subadditivity.

\subsec{Entanglement Entropy for Straight Belt}

Before we move on to the holographic dual description, we would like
to mention the simplest example of entanglement entropy. This is the
case where the submanifold $A$ is given by the $d$ dimensional
straight belt $A_S$ (refer also to \RT \RTL). It has the width $l$
in one direction and extends in other $d-1$ directions infinitely
(its regularized length is denoted by $L$). In general, the
entanglement entropy $S_A$ in $d+1\geq 3$ dimensional CFT takes the
following form \eqn\generalform{S_A=\gamma\cdot
\f{L^{d-1}}{a^{d-1}}-\beta\cdot \f{L^{d-1}}{l^{d-1}},} where
$\gamma$ and $\beta$ are positive constants. This form \generalform\
has been obtained from the explicit computation in free field
theories \casinifree\ and holographic results \RT \RTL. The $d=1$
case is special and the entropy is known to be expressed as \Cardy\
\eqn\tden{S_{A}=\f{c}{3}\log (l/a),} where $c$ is the central
charge.

Following discussions similar to the ones in the previous
subsections, we can find that the strong subadditivity requires that
the finite term $S_{finite}(l)$ of the entropy satisfies
$\f{d^2}{dl^2}S_{finite}\leq 0$. Indeed, this is clearly true for
\generalform\ and \tden.

Moreover, we can obtain a stronger condition\foot{The derivation of
this stronger condition relies heavily on the simple structure of
the submanifold $A$ and cannot be directly applied to other examples
discussed in section 3.1 and 3.2.} from a relativistic consideration
of the strong subadditivity as discovered in \Casinithree. This
requires
\eqn\strongerc{\f{d}{dl}\left(l\f{d}{dl}S_{finite}\right)\leq 0.}
Again this is satisfied by \generalform\  and \tden.

In two dimensional CFTs, \strongerc\ leads to the entropic c-theorem
\Casinifour. Define the entropic $c$-function $C$ by
\eqn\entroc{l\f{dS_A}{dl}=C(l),} in non-conformal field theories.
$C(l)$ approaches $c/3$ at a fixed point. Then \strongerc\ is
equivalent to the c-theorem $C'(l)\leq 0$ \Casinifour, regarding
naturally $l$ as the length scale. It will be a very exciting future
problem to extend this entropic proof of c-theorem to higher
dimensions, though it does not seem to be straightforward.

\newsec{Holographic Entanglement Entropy and Bousso Bound}

\subsec{Holographic Entanglement Entropy}

The holographic principle tells us that the true degree of freedom
in a $d+2$ dimensional gravity theory is actually $d+1$ dimensional
\susskind. This idea is manifestly realized in string theory as the
AdS/CFT correspondence \Mal. In principle, we believe that we can
compute any physical quantities in a $d+1$ dimensional CFT from the
dual $d+2$ dimensional anti de-Sitter space ($AdS_{d+2}$). Thus one
direct way to obtain a gravitational interpretation of the
entanglement entropy in a CFT is to apply the AdS/CFT
correspondence.

In \RT \RTL , it is claimed that the entanglement entropy can be
computed as follows \eqn\minien{S_A=\f{A(\gamma_A)}{4G^{(d+2)}_N},}
where $A(\Sigma)$ denotes the area of the surface $\Sigma$, and
$G^{(d+2)}_N$ is the Newton constant in the $d+2$ dimensional anti
de-Sitter space. The $d$ dimensional surface $\gamma_A$ is
determined such that the minimal area surface whose boundary
coincides with the boundary of submanifold $A$.

Intuitively, \minien\ can be understood by applying the
Bekenstein-Hawking entropy formula to the surface $\gamma_A$ as if
it is an event horizon as we expect that $\gamma_A$ represents the
lost information hidden inside the region $B$ in gravity
description. As checked in \RT, the formula \minien\ is exactly true
in $d=1$ examples. We can also find partial quantitative evidences
in higher dimensional cases \RTL.

Moreover, as shown in \Fur\ recently, we can derive \minien\ from
the basic principle of AdS/CFT correspondence \GKP. Its outline can
be summarized very briefly as follows. Consider the partition
function $Z_{CFT}$ of a CFT in the presence of the negative deficit
angle $2\pi-\delta<0$. It is the same as the trace $\tr_A\ \rho_A^n$
by setting $n=\f{\delta}{2\pi}$. The locus of the deficit angle is
localized on a codimension two surface which is identified with the
boundary $\de A$. Then the entanglement entropy in the CFT is equal
to the derivative of partition function with respect to the deficit
angle (refer to e.g. \RTL) \eqn\derivativepar{S_A=-\f{\de}{\de n}
\log \tr_A\ \rho_A^n|_{n=1}=-2\pi \f{\de}{\de\delta}\log
Z_{CFT}|_{\delta=2\pi}.}

 On the other hand, the partition function in
CFT$_{d+1}$ can be obtained as that of the supergravity on the
AdS$_{d+2}$ space \GKP\ via the AdS/CFT. At the tree level in the
supergravity we thus find  $Z_{CFT}=e^{-S_{sugra}}$, where
$S_{sugra}$ is the tree level supergravity action. The deficit angle
at the boundary $z=a$ is extended into the bulk of the AdS space and
will form a $d$ dimensional surface $\gamma_A$. Then the
Einstein-Hilbert term $\f{1}{16\pi G_N^{(d+2)}}\int dx^{d+2} \s{G}R$
is evaluated as $\f{A(\gamma_A)}{8\pi G_N^{(d+2)}}(\delta-2\pi)$
since the scalar curvature behaves as a delta function localized at
the surface $\de A$. Finally, we apply the ordinary variational
principle we find that the surface $\gamma_A$ should be the minimal
surface and it gives the largest contribution to the path-integral
of the gravity partition function. In this way we recover \minien\
using \derivativepar.

\subsec{Relation to the Bousso Bound}

To apply the Bekenstein-Hawking formula to a surface which is
actually not an event horizon, looks similar to the idea of the
entropy bound \susskind \Bousso. Motivated by this, in this
subsection we would like to discuss the relation between
entanglement entropy and covariant entropy bound, which is known as
the Bousso bound \Bousso.

The Bousso bound is the following claim. Consider a space-like
codimension two manifold $\Sigma$ with the area $A(\Sigma)$. Its
light-sheet $L(\Sigma)$ is defined to be a codimension one
hypersurface bounded by $\Sigma$ and generated by one of the four
null congruence orthogonal to $\Sigma$. We choose the light-sheet
such that the expansion of light rays is always negative. Let
$S_{L(\Sigma)}$ be the entropy on the light-sheet. Then the Bousso
bound argues \eqn\bousso{S_{L(\Sigma)}\leq \f{A(\Sigma)}{4G_N}.} We
would like to apply this bound to the spacetime $AdS_{d+2}$.

Now we go back to the entanglement entropy in CFTs. Physically the
entanglement entropy $S_A$ measures the loss of information when we
smear out the region $B$. As we have explained, $S_A$ is specified
by the Cauchy horizon $H^+(B)$. Thus its bulk description should be
such that the light-sheet, where the amount of information (or
entropy) is measured, ends on $H^+(B)$ at the boundary. Since
$L(\Sigma)$ is defined such that $L(\Sigma)\perp \Sigma$,
 the spacelike manifold $\Sigma$ is
orthogonal to the boundary when they meet. This is always satisfied
if we choose $\Sigma$ the minimal surface $\gamma_{A}$. Notice that,
however, we can choose other surface with this property. In this way
we find the setup in Fig.4.

\fig{Relation to Bousso bound. We write the light-sheet in the
Poincare coordinate $ds^2=z^{-2}(-dt^2+dx^2+dz^2)$ of $AdS_3$.
Notice the form of light-cone is the same as that in the flat
spacetime. The light-sheet is given by a half cone $L$ and the
minimal surface $\gamma_A$ is a half
circle.}{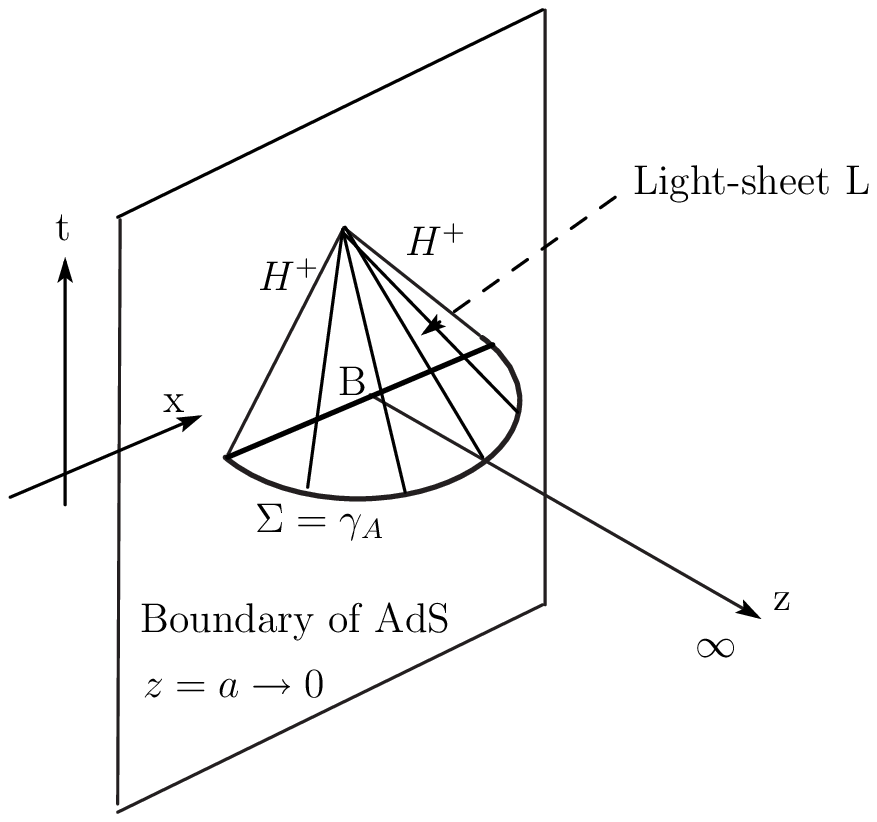}{3truein}

Now we apply the Bousso bound to this system. The entropy on the
light-sheet is bounded by the area of $\Sigma$. If $\Sigma$ is
chosen to be the minimal surface $\gamma_A$, we find
$S_{L(\gamma_A)}\leq \f{A(\gamma_A)}{4G_N}$ from \bousso. Following
the general idea of holography we would like to claim that the
entanglement entropy is the same as the entropy on the light-sheet
$L(\gamma_A)$. Then the relation \minien\  claims that {\it the
entanglement entropy $S_A$ saturates the Bousso bound}\ \bousso
\eqn\bsa{ S_A=S_{L(\gamma_A)}=\f{A(\gamma_A)}{4G_N}.} All of these
claims can be shown to be exactly true in the $AdS_3/CFT_2$ case as
the entanglement entropies in 2D CFTs are analytically computable
\RT. Also notice that even if we do not choose $\gamma_A$ as
$\Sigma$, the previous condition that $\Sigma$ should be orthogonal
to the boundary is enough to prove that the leading divergent term
in $\f{A(\Sigma)}{4G_N}$ is the same as the leading (area law)
term\foot{It is clear that these arguments also hold for
asymptotically AdS spaces. The dual theories are quantum field
theories with UV fixed points. One of the examples which do not
satisfy this criterion is the 4D cascading theory \KS. In this
model, The radius of the five dimensional Einstein manifold $R$
depends on the energy scale or equally the radial coordinate $r$ as
$R\sim (\log r)^\f14$ in the UV region. This corresponds to the fact
that the rank of gauge groups increases at higher energy in the
cascading gauge theory. The calculation of minimal surface areas can
be done as before and in the end we find that the leading divergence
is proportional to $ -\f{|\de A|}{a^{2}}\cdot (\log a)^2$. Thus in
this case, the well-known area law of the leading divergence
$\sim\f{|\de A|}{a^{d-1}}$ (see also \generalform ) is modified by
the logarithmic factor.} in the holographic entanglement entropy
\minien.

We also need to explain why the bound is saturated. One explanation
is clearly to remember the proof found in \Fur\ and to directly
compute the entropy following the basic principle of AdS/CFT \GKP.
Even if we do not apply the AdS/CFT, we can at least tell that this
is strongly suggested. This is because we are choosing the minimal
surfaces among other candidates of $\Sigma$. If we employ any one of
the others, we will have a weaker bound since $A(\gamma_A)<
A(\Sigma)$ for all $\Sigma$s other than $\gamma_A$. Thus the minimal
surface offers the strictest bound in \bousso\ and thus there is an
opportunity to saturate the bound.

These arguments clarifies the following important question. In which
region of the $AdS$ space is encoded the information in the dual
CFT? Our example depicted in Fig.4 manifestly shows that the
information in the $B$ region, which is equal to the one in its
causal development $D^+(B)$, is dual to the inside of the
light-sheet $L$.

\newsec{Holographic Evidences for Strong Subadditivity}

In this section we would like to discuss the main issue of this
paper.  This is to examine if the strong subadditivity is satisfied
in the holographic dual computations. For this purpose, we will
study several particular examples of entanglement entropy in $2+1$
and $3+1$ dimensional CFTs. In the dual $AdS_{d+2}$ spacetime, we
always employ the Poincare coordinate
\eqn\poincarecor{ds^2=R^2\f{dz^2-dx_0^2+\sum_{i=1}^{d}dx_i^2}{z^2}.}
The radial coordinate $z$ represents the length scale and we put the
UV cutoff at $z=a$. To compute the entanglement entropy, we choose a
particular time slice e.g. $x_0=0$ and then try to find the minimal
surface $\gamma_A$. We require that the boundary of $\gamma_A$ at
$z=a$ coincides with the boundary of the submanifold $A$ (or equally
its complement $B$) which defines the entanglement entropy $S_A$ as
explained in section 4. Finally we apply the formula \minien\ to
obtain the entropy $S_A$. Notice that this is generally divergent in
the limit $a\to 0$ and the leading divergence takes the form $\sim
\f{|\de A|}{a^{d-1}}$, which agrees with the known area law of the
entanglement entropy.

The results found for $(2+1)$ dimensional CFTs via holographic
computations can be equally applied to the calculation of Wilson
loops in $(3+1)$ dimensional CFTs which are defined by strongly
coupled gauge theories. Therefore we can claim that the strong
subadditivity relation is also true for the Wilson loop correlation
functions. This issue will be discussed briefly in the section 6.

\subsec{Cusp Case}

First we would like to discuss the $d=2$ example (i.e. (2+1) dim.
CFT) where the subsystem $A$ has a cusp singularity. This is the
same setup as the one discussed in section 3.1. The boundary $\de A$
is a line with a cusp at a point as in Fig.2. Its angle is defined
to be $\Omega$. In the polar coordinate $(r,\phi)$ of the
$(x_1,x_2)$-plane, it is just described by
\eqn\cusp{\{(r,\phi)|0\leq r<\infty,\phi=0\}\cup \{(r,\phi)|0\leq
r<\infty,\phi=\Omega\}.}

Next we would like to find the corresponding minimal surface
$\gamma_A$ in $AdS_{d+2}$. Fortunately, this setup is essentially
the same as that appears in the Wilson loop calculation \Wilson\
done in \DGO\ and thus we can employ the results there.

Using the conformal symmetry $z\to \lambda z, x_\mu \to \lambda
x_\mu$, we can assume that the surface $\gamma_A$ is described \DGO\
by \eqn\gammaa{z(r,\phi)=\f{r}{g(\phi)}.}

The area of $\gamma_{A}$ is given by \eqn\aregama{|\gamma_A|=
R^2\int \f{dr}{r}\int d\phi \s{(g')^2+g^2+g^4}.} Thinking $\phi$ as
a time, we find the Hamiltonian  \eqn\ennrgy{H=\f{\de L}{\de
g'}g'-L= -\f{g^2+g^4}{\s{(g')^2+g^2+g^4}}=const.(\equiv
-g_0^2-g_0^4).}

Then we can reduce the minimal surface equation to the following
differential equation \eqn\minimals{\f{dg}{d\phi}=g\s{(1+g^2)\left(
\f{g^2+g^4}{g_0^2+g_0^4}-1\right)}.}

Since $g=g_0$ is the turning point, we have to require
\eqn\condfi{\eqalign{\f{\Omega}{2}&=\int^{\infty}_{g_0}\f{dg}{g\s{(1+g^2)
\left(\f{g^2+g^4}{g_0^2+g_0^4}-1\right)}}\cr
&=g_0\s{1+g_0^2}\int^\infty_0\f{dz}{(z^2+g_0^2)\s{(z^2+g_0^2+1)(z^2+2g_0^2+1)}}
,}} where we set $g^2=z^2+g_0^2$.

In the end the minimal area can be found as
\eqn\areacusp{\eqalign{|\gamma_A|&=2R^2\int^{\infty}_{g_0a}
\f{dr}{r}\int^{r/a}_{g_0}dg\f{g\s{1+g^2}}{\s{(g^2-g_0^2)(g^2+g_0^2+1)}}\cr
&=2R^2\int^{\infty}_{g_0a} \f{dr}{r} \int^{r/a}_0
dz\s{\f{z^2+g_0^2+1}{z^2+2g_0^2+1}},}} where we noticed that the
upper bound of the $g$ integral is $r/a$ from \gammaa. The
integration of $g$ diverges as $\f{r}{a}-f(\Omega)$ when the cutoff
$a$ is set to zero. The finite part is explicitly given by
\eqn\fini{f(\Omega)=\int^{\infty}_0
dz\left[1-\s{\f{z^2+g_0^2+1}{z^2+2g_0^2+1}}\right],} where the
relation between $\Omega$ and $g_0$ is determined from \condfi. Then
the total area can be found as (the upper bound of $r$ is defined to
be $L$ for a regularization)
\eqn\totalarea{\f{|\gamma_A|}{R^2}=\f{2L}{a}-2f(\Omega)\log\f{L}{a}
+(\rm{finite\ terms}).} Thus the entanglement entropy is now
computed as (we omit finite constant terms)
\eqn\entan{S_A=\f{R^2}{4G^{(4)}_N}
\left(\f{2L}{a}-2f(\Omega)\log\f{L}{a}\right).} Note that the
function $f(\Omega)$ is the same as the one in \onetwologdim\ up to
the factor $\f{R^2}{2G_{N}^{(4)}}$.

The strong subadditivity tells us that the function $f(\Omega)$ is a
convex function as we have seen in section 3.1. Indeed we can check
this property from the explicit form of $f(\Omega)$ shown in Fig.5.
\fig{The function $f(\Omega)$ in $(2+1)$D CFT; the horizontal and
vertical coordinate are given by $\Omega/\pi$ and $f(\Omega)$,
respectively. $f(\Omega)$ becomes vanishing at
$\Omega=\pi$.}{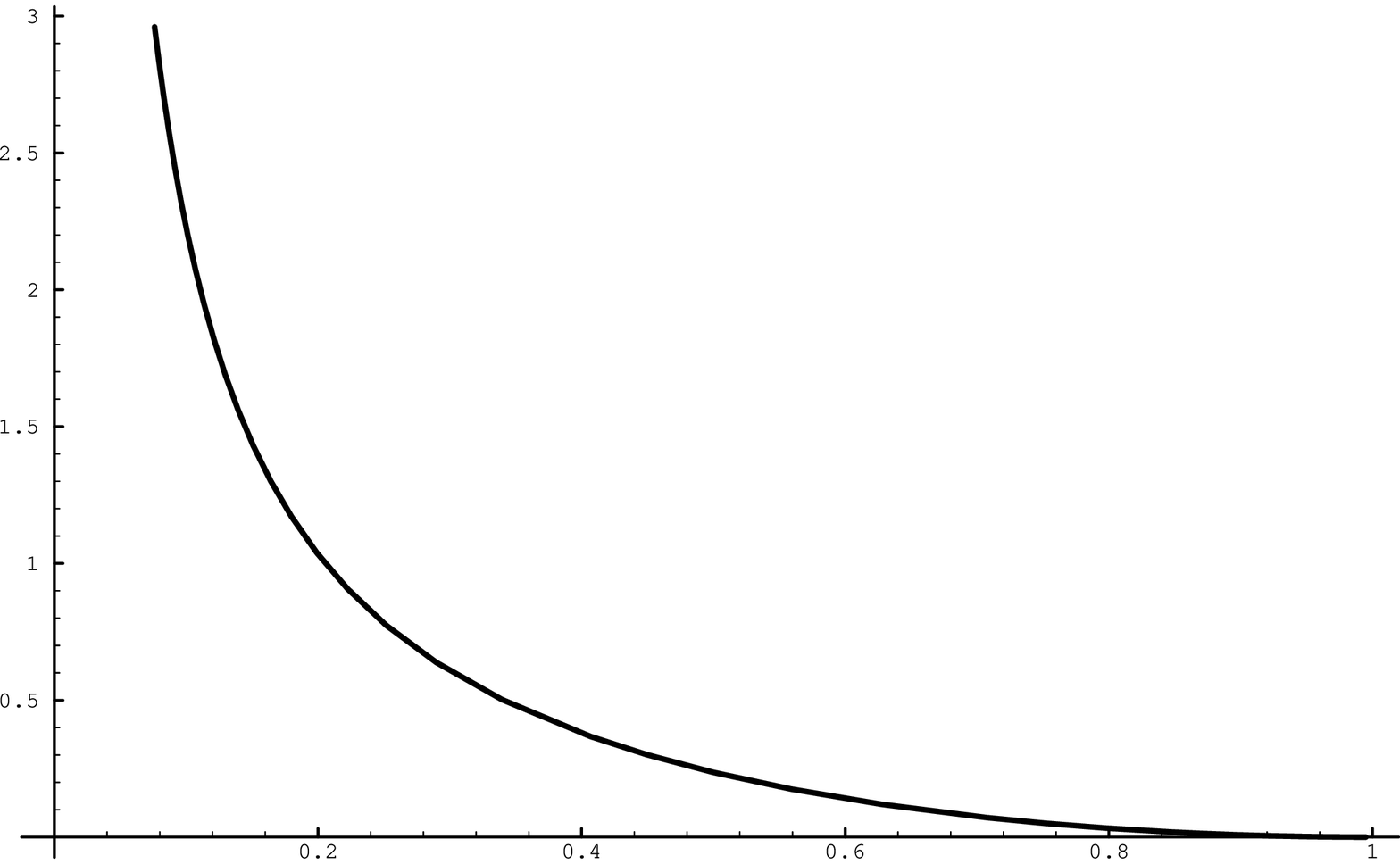}{3truein}

It is also possible to find the following analytic form of
$f(\Omega)$ when $\Omega$ is very small
\eqn\analyticf{f(\Omega)\simeq
\f{\Gamma\left(\f{3}{4}\right)^4}{\pi}\cdot \f{1}{\Omega}.} This
behavior $f(\Omega)\propto \Omega^{-1}$ for a small angle is the
same as the one computed in the free scalar field theory\foot{ In
the free field theory, the coefficient $\beta$ in front of
$\Omega^{-1}$ in the logarithmic term of $S_A$
($\beta=\f{R^2}{2G^{(4)}_N}\cdot\f{\Gamma\left(\f{3}{4}\right)^4}{\pi}$
in our case of \analyticf) is equal to $\f{1}{\pi}\int^\infty_0 dt
c(t)$, where $c(t)$ is the entropic c-function \CHcusp. Also in the
same paper \CHcusp\ it is argued and confirmed in the free field
theoretic computations that this quantity $\beta$ is equal to the
coefficient $\beta$ of the entropy in the straight line case (see
\generalform). We can check that this relation is indeed true in our
case exactly. We are very grateful to H. Casini for explaining this
relation to us. } \CHcusp.

Before we jump to the next example, it may be helpful to apply our
formula \entan\ to a specific background. Consider the $AdS_4\times
S^7$ realized as the near horizon limit of $N$ M2-branes in M-theory
. The corresponding CFT is defined by the strong coupling limit of
the (2+1) dimensional maximally supersymmetric gauge theory. The
entropy is explicitly given by
\eqn\erntromtwo{S_A=\f{N^\f32}{3\s{2}}\left(\f{2L}{a}-2f(\Omega)\log\f{L}{a}\right).}

\subsec{Annulus Case}

Next we examine the case where the subsystem $A$ is given by the
annulus whose boundary $\de A$ consists of two concentric rings
 with the radius $r_1$ and $r_2$ (assume $r_1<r_2$). We again use the polar
 coordinate for the $(x_1,x_2)$-plane. We can also consider higher
 dimensional cases where the concentric rings are replaced by concentric $d-1$
 dimensional spheres. These are the same setups as the ones discussed in section
 3.2. Below we will mainly be interested in the $d=2$
 and  $d=3$ case (i.e. (2+1) and (3+1) dimensional CFTs).

The minimal surface is the $d$ dimensional `half torus' defined by
the half circle $z=z(r)$ times the sphere $S^{d-1}$ s.t.
$z(r_1)=z(r_2)=a\to 0$. The function $z(r)$ is found by minimizing
the area functional \eqn\minimalena{|\gamma_A|=R^d \cdot
{\rm{Vol}}({\rm{S}}^{d-1}) \cdot\int^{r_2}_{r_1} dr
r^{d-1}\f{\s{1+(\f{dz}{dr})^2}}{z^d}.}  We can find the following
equation of motion
\eqn\eoma{rzz''+(d-1)z(z')^3+(d-1)zz'+dr(z')^2+dr=0.} Its simplest
solution is the $d$ dimensional sphere $z^2+r^2=$const. Clearly we
have to deal with more complicated solutions to find the desired
configurations. Thus one way to analyze this problem is to directly
resort a numerical analysis as explained below.

First let us concentrate on the $d=2$ case. The numerical
investigation shows that there is an upper limit on the allowed
value of $r_2/r_1$. This means that if $r_2/r_1$ is large enough,
then there is no solution to \minimalena. In this case the minimal
area surface should be regarded as the two disconnected spheres with
the radii $r_1$ and $r_2$. This critical value can be estimated as
$(r_2/r_1)_*\simeq 2.725$. This kind of phase transition has already
been known in the context of Wilson loop computations and is called
Gross-Ooguri transition \GO \Za \Kim. We solved \eoma\ numerically
and calculated the finite part of the integral (setting $\rho=\log
(r_2/r_1)$) \eqn\finib{\ti{b}(\rho)=\left[\int^{r_2}_{r_1} dr
r\f{\s{1+(\f{dz}{dr})^2}}{z^2}\right]-\f{r_1+r_2}{a}.} The result is
plotted in Fig.6. The total entanglement entropy is expressed as
\eqn\entangb{S_A=\f{|\gamma_A|}{4G^{(4)}_N} =\f{2\pi
R^2}{4G^{(4)}_N}\left(\f{r_1+r_2}{a}+\ti{b}(\rho)\right).} The
entanglement entropy\foot{We can also apply the above result for
$d=2$ to the Wilson loop computation. This leads to the following
correlation function of two concentric Wilson loop operators in the
strongly coupled large $N$ ${\cal N}=4$ super Yang-Mills theory $\la
W(r) W(r')\lb^{\rm{connected}}
=e^{-\s{2g_{YM}^2N}\cdot\ti{b}(\rho)}. $} in the specific example of
$AdS_4\times S^7$ is obtained by substituting
$\f{R^2}{4G^{(4)}_N}=\f{N^\f32}{3\s{2}}$.

Now we can see that there are two branches of solutions when
$r_2/r_1<(r_2/r_1)_*$ as we showed in Fig.6. One of them is
uninteresting since its area is larger than the trivial one (i.e.
two disconnected spheres $\ti{b}_{sphere}=-2$). The second one is
physically relevant as long as $\ti{b}<-2$ (i.e.$\rho<0.88$). Above
that point, the trivial one becomes more stable and thus we have
$\ti{b}(\rho)=-2$.

In summary, the physically relevant minimal surface is found as
follows. When $\ti{b}<  \ti{b}_{sphere}$, it is given by one of the
two non-trivial connected solutions to \eoma\ with the smaller area.
At $\ti{b}=  \ti{b}_{sphere}$, a transition takes place. Then it
 becomes the disconnected
sphere solution when $\ti{b}>  \ti{b}_{sphere}$. Therefore we can
check from Fig.6 that the finite part $b(\rho)=\f{2\pi
R^2}{4G^{(4)}_N}\ti{b}(\rho)$ of the entanglement entropy is a
concave function of $\rho$ (i.e. $b''\leq 0$) and this agrees with
the strong subadditivity discussed in section 3.2. It may also be
intriguing to note that the first irrelevant solution does not
satisfy the concave condition. For that solution, the value of
$\ti{b}$ approaches $\ti{b}_{sphere}=-2$ in the $\rho\to 0$ limit.

\fig{The function $\ti{b}(\rho)$ computed from the minimal surfaces
in $AdS_4$ ($d=2$). The horizontal and vertical coordinate are
 $\rho$ and $\ti{b}$, respectively.  The dotted points are obtained
from the direct numerical solutions to \eoma , while the solid
curves are from the analytical method discussed in (5.19). Both of
them agree with each other very well. There are two solutions and
the lower one is physically relevant as it has a smaller
area.}{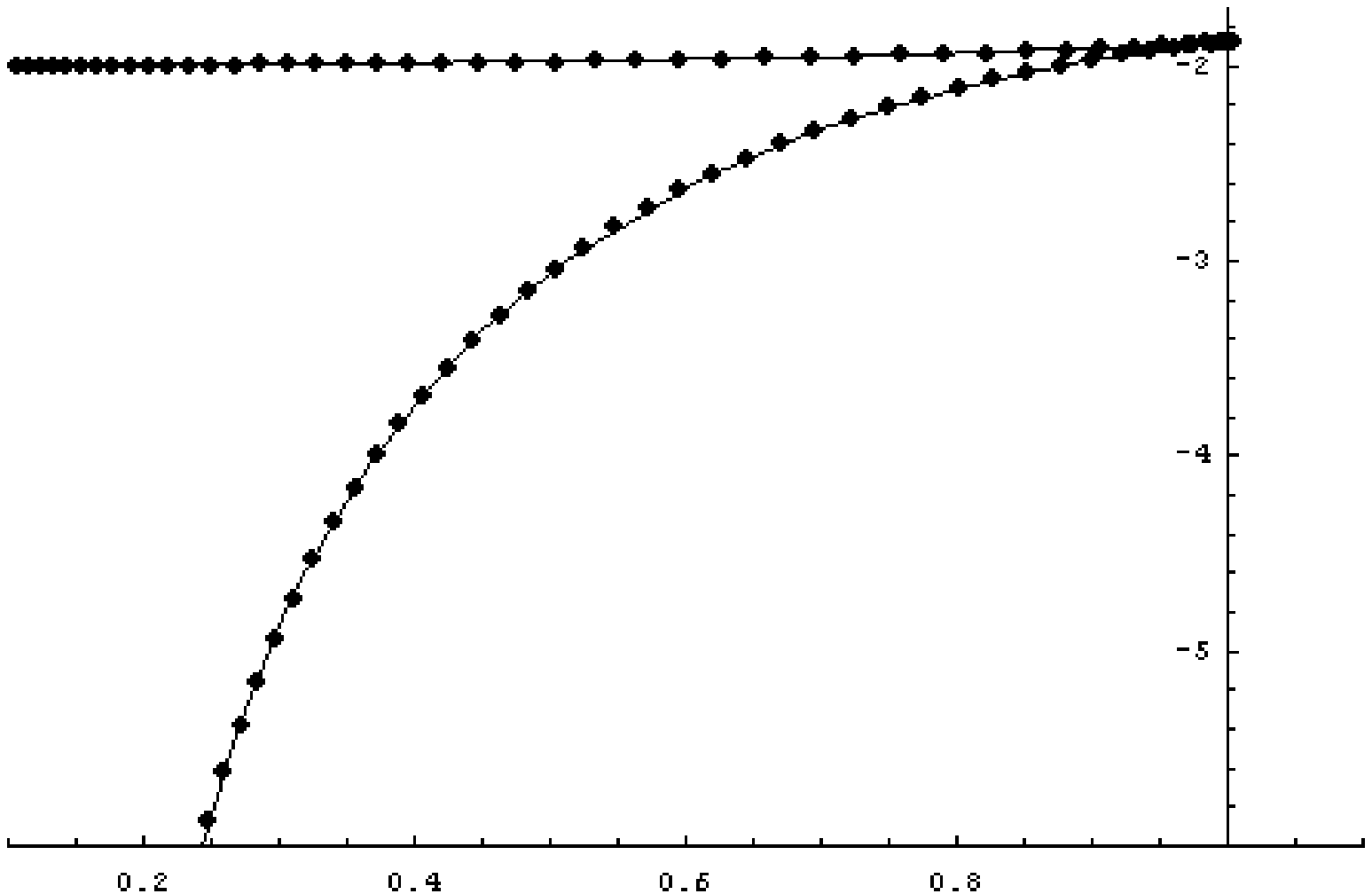}{3truein}

\vskip .2in

Now we turn to the $d=3$ case. The entropy takes the following form
\eqn\entrodth{S_A=\left(\f{R^3}{4G^{(5)}_N}\right)\cdot
4\pi\left[\f{r^2+r'^2}{2a^2}
-\f{1}{2}\log\left(\f{rr'}{a^2}\right)+\ti{b}_3\left(\f{r'}{r}\right)\right].}
In the specific case of $AdS_5\times S^5$ dual to the ${\cal N}=4$
$SU(N)$ super Yang-Mills, the entropy is given by substituting
$\f{R^3}{4G^{(5)}_N}=\f{N^2}{2\pi}$ in \entrodth. The coefficient of
leading divergence and logarithmic divergence in \entrodth\ can be
fixed by noting that when $z$ is small, the solution to \eoma\ is
well approximated by the simple sphere solutions\foot{This fact can
be checked by looking at the asymptotic behavior at $z=0$ of the
function $r=r(z)$, which satisfies the differential equation
$zrr''-(d-1)(r')^2z-(d-1)z-drr'-dr(r')^3=0$ equivalent to \eoma.}.

\fig{The function $\ti{b}_3(\rho)$ computed from the minimal
surfaces in $AdS_5$ ($d=3$). The horizontal and vertical coordinate
are $\rho$ and $\ti{b}$, respectively.  The dotted points are
obtained from the direct numerical solutions to \eoma , while the
solid curves are from the analytical method discussed in (5.21).
Both of them agree with each other very well. There are two
solutions and the lower one is physically relevant as it has a
smaller area.}{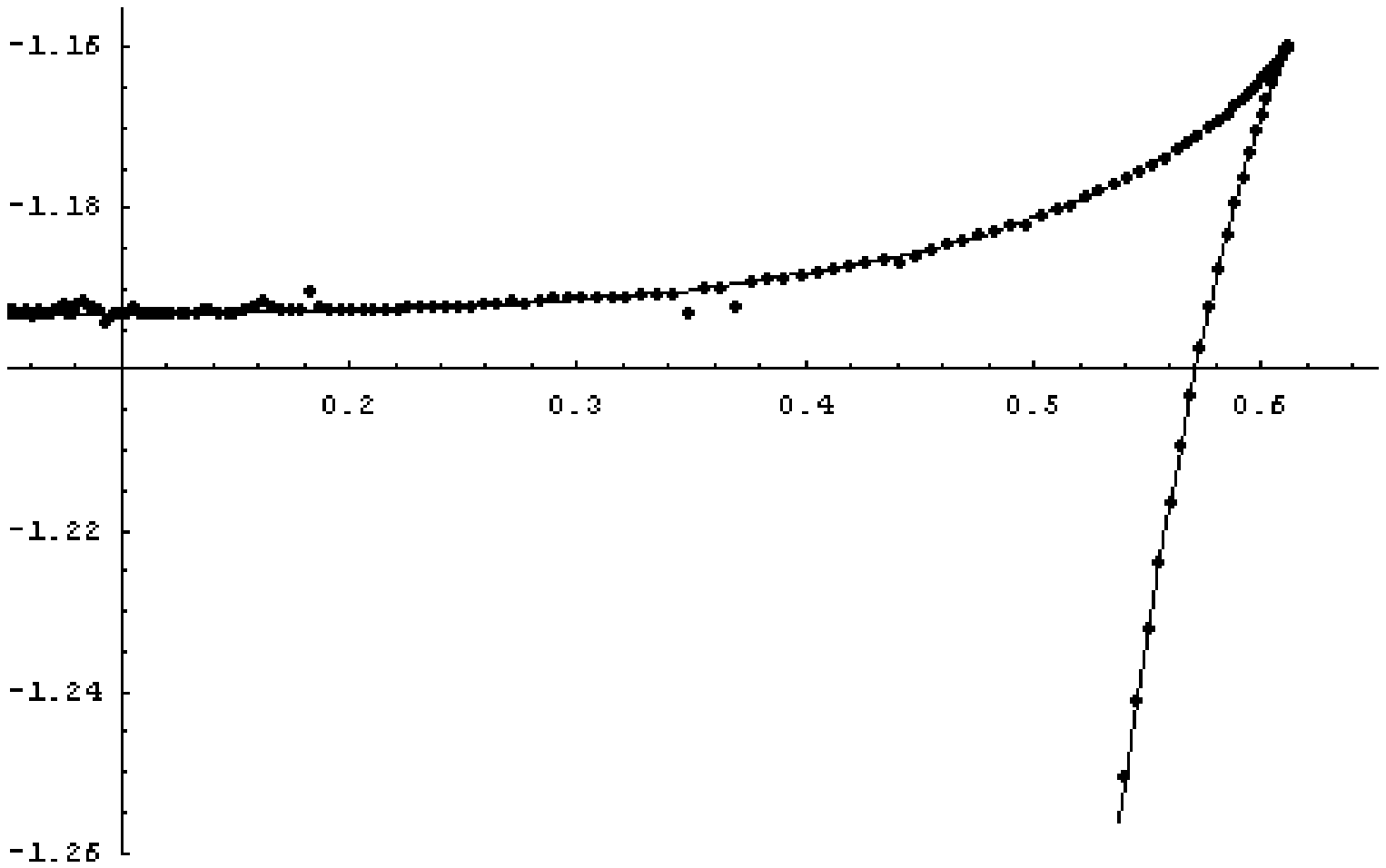}{3truein}

We again solved \eoma\ numerically and computed the finite term
$\ti{b}_3$. This is plotted in Fig.7. The properties of the
solutions to \eoma\ at $d=3$ are similar to the previous one at
$d=2$. Again two different solutions exist when
$r_2/r_1<(r_2/r_1)_*=1.844$. Above this value there is no connected
solution. The value of $\ti{b}_3$ for one of the two solutions is
always greater than that of two disconnected spheres
$\ti{b}_{3sphere}=-1/2-\log2 \simeq -1.193$ and it approaches this
value $\ti{b}_{3sphere}$ in the $\rho\to 0$ limit. Thus this
solution is physically irrelevant.

In summary, the physically relevant minimal surface is found as
follows. When $\ti{b}_3< \ti{b}_{3sphere}$, it is given by one of
the two non-trivial connected solutions to \eoma\ with the smaller
area. At $\ti{b}_3= \ti{b}_{3sphere}$, a transition takes place.
Then it becomes the disconnected sphere solution when $\ti{b}_3>
\ti{b}_{3sphere}$. Thus again
 we can confirm that this function $\ti{b}_3(\rho)$ is concave,
 agreeing with the
strong subadditivity.

\vskip .2in

Finally we would like to briefly mention a semi-analytical approach
to this problem. Let us artificially add an extra coordinate
$x^{d+1}$ in \poincarecor\ (i.e. $AdS_{d+3}$) and consider the $d$
dimensional minimal surface whose boundary is given by two
concentric rings (or spheres). We first assume that they are
separated from each other by $\Delta x^{d+1}$ in the new $x^{d+1}$
direction. When $d=2$, this example has been studied in \Za \Kim\ in
order to compute the Wilson loops. In this setup, we can find first
integrals and reduce the problem to the analysis of first order
differential equations\foot{After we submitted this paper, we
noticed that in the recent paper \DF , the explicit solution at
$\Delta x^{d+1}=0$ was obtained in the context of Wilson loop. We
are grateful to N. Drukker for letting us know the paper.}
 \Za \Kim\ if we treat all quantities as
functions of $x^{d+1}$.

Further, we take the zero separation limit $\Delta x^{d+1}\to 0$.
Then we go back to the original setup discussed in the above to
compute the entanglement entropy (i.e. \eoma ). However, we have to
be careful since the radii of the rings become zero in this limit.
If we neglect this point, we can find the integral expression of the
area $|\gamma_A|$. These have already been expressed in Fig.6 and
Fig.7 as a solid curve. Indeed, we can see that they nicely agree
with the results obtained by solving \eoma\ numerically. Thus we can
find that this method offers us the exact computations of $\ti{b}$
and $\ti{b}_3$.

Their explicit forms are summarized as follows. In the $d=2$ case,
we first define the parameter $s$ in terms of $\rho(\equiv\log
(r_2/r_1))$ as follows \eqn\exptww{\rho=s\int^{\sin^2\theta_0}_0
dy\f{y^{1/2}}{\s{(1-y)(1-y-s^2 y^2)}},} where $\theta_0$ is the
solution to $s^2 \sin^4\theta_0=\cos^2\theta_0$. Then the area
becomes \eqn\areath{\f{|\gamma_A|_{d=2}}{2\pi R^2}
=\int^{\theta_0}_{a/r_1}d\theta\f{\cos\theta^2}
{\sin\theta^2\s{\cos^2\theta
-s^2\sin^4\theta}}+\int^{\theta_0}_{a/r_2}
d\theta\f{\cos\theta^2}{\sin\theta^2\s{\cos^2\theta
-s^2\sin^4\theta}}.} In the $d=3$ case, we define
\eqn\expdth{\rho=s\int^{\sin^2\theta_0}_0
dy\f{y}{\s{(1-y)\left((1-y)^2-s^2 y^3\right)}},} where $\theta_0$ is
the solution to $s^2 \sin^6\theta_0=\cos\theta_0^4$. Finally the
area becomes \eqn\areaths{\f{|\gamma_A|_{d=3}}{4\pi R^3}
=\int^{\theta_0}_{a/r_1}d\theta\f{\cos\theta^4}{\sin\theta^3\s{\cos^4\theta
-s^2\sin^6\theta}}+\int^{\theta_0}_{a/r_2}
d\theta\f{\cos\theta^4}{\sin\theta^3\s{\cos^4\theta
-s^2\sin^6\theta}}.} The functions $\ti{b}(\rho)$ and
$\ti{b}_3(\rho)$ can be obtained from \areath\ and \areaths\ by
subtracting the divergences.

\newsec{Conclusions and Discussions}

In this paper, we explored further evidences for the recent proposal
of holographic computations of entanglement entropy. Especially we
looked at an important property of the entanglement entropy, known
as the strong subadditivity.

First we showed that this requires various terms in the entanglement
entropy should be concave functions with respect to the geometric
parameters such as the cusp angle or the ratio of radii of an
annulus. Secondly we computed the entanglement entropy from the dual
gravity side in several explicit examples and found that the strong
subadditivity is satisfied in all of our examples. This offers us a
highly non-trivial check of our proposal. It will be obviously an
important future problem to derive the required concavity of the
minimal surface area in AdS spaces from a more systematic method,
which probably needs the use of a positive energy condition and of
the Einstein equation on AdS spaces.

Our results in $(2+1)$D CFT can directly be applied to the Wilson
loop computations in 4D gauge theories via the AdS/CFT
correspondence. Thus we are led to the conjecture that the strong
subadditivity relation \strongtwo\ is also true for the (locally
supersymmetric) Wilson loops $\la W(C)\lb$ by replacing $S_A$ with
$-\log\la W(\de A) \lb$ at least in strongly coupled gauge theories.
Indeed, the concavity of the quark-antiquark potential has been
already shown in \Ba\ , which corresponds to the particular case of
rectangular Wilson loops. This property has been successfully tested
in the AdS/CFT correspondence in \GOW \DP\ for some specific
examples. This issue clearly deserves further investigations.

In addition, we discussed the relation between the holographic
entanglement entropy and the covariant entropy bound known as the
Bousso bound. We noted that the holographic entanglement entropy
formula \minien\ implies that the entropy bound is saturated. This
leads to another natural explanation of the proposal. At the same
time, this makes clear the region in AdS space where are encoded the
information included in a specified submanifold of the boundary CFT.

\vskip .5in

\centerline{\bf Acknowledgments} We are extremely grateful to H.
Casini and S. Ryu for the careful reading of the manuscript and for
valuable discussions. We thanks A. Shirasaka and S. Yamato for
useful discussions, T. Azeyanagi for careful reading and useful
comments, and T. Muranushi for technical supports on numerical
computations. We would also like to thank N. Drukker for letting us
know relevant and useful papers. The research of TH was supported in
part by JSPS Research Fellowships for Young Scientists.

\vskip .3in

\listrefs

\end